# Solar Cycle 25 Dynamics from Observational and Statistical Parameters Characterization of the Maximum Phase and Rotational Behaviour

Homer Dávila Gutiérrez[1,2]

[1] SKYCR.ORG, San José, Costa Rica
[2] Master in Astrophysics, Universidad Internacional de La Rioja (UNIR), 26006 Logroño, Spain
ORCID: https://orcid.org/0009-0002-5968-3646
E-mail: cosmos@skycr.org

**Abstract**
We present a comprehensive analysis of Solar Cycle 25 aimed at precisely constraining the interval of its activity maximum using multiple observational parameters: sunspot number (SSN), Wolf number, the 10.7 cm solar radio flux (F10.7), the occurrence rate and speed of coronal mass ejections (CMEs), the statistics of X-ray flares, and the reversal of the global magnetic field. A percentile-based classification of the adjusted F10.7 flux (EPH scale) is introduced, which allows us to identify the maximum-activity interval as August 2024 to January 2025. This period coincides with the reversal of the solar polar fields and is characterized by enhanced eruptive activity, including more than 190 M-class and 19 X-class flares (among them an X6.3 flare on 10 December 2024) and multiple CMEs with speeds exceeding 2000 km s$^{-1}$. In the second part of the study we investigate solar rotation during this cycle using an independent observing campaign conducted in April 2025 in San José, Costa Rica, with privately owned telescopes in white light and in the Hα line. Photometric techniques and active-region tracking yield mean synodic angular velocities of 12.8 ± 2.8° day$^{-1}$ in the northern hemisphere and 13.38 ± 0.54° day$^{-1}$ in the southern hemisphere, corresponding to sidereal periods of 22.78 and 21.77 days, respectively. Complementary SWPC–NOAA sunspot-region data provide synodic periods between 21.19 and 25.74 days, consistent with classical differential-rotation measurements and confirming the expected latitudinal gradient. Linear velocities derived from SDO images span from about 872 km h$^{-1}$ to more than 7400 km h$^{-1}$, depending on latitude and date.

### 1. Introduction

The Sun is a fundamental laboratory for the study of plasma physics, magnetohydrodynamics, and star–planet interaction (Athay 1976; Beckers 1964; Rutten 2003; Priest 2014). Its magnetic activity modulates space weather, affecting the ionosphere, communications, navigation, and technological infrastructure on Earth (Tapping 2013; Chen 2011).

Solar activity is organized in cycles of approximately 11 years, characterized by variations in sunspot number, F10.7 flux, and eruptive activity. Two ≈ 11-year cycles constitute a complete magnetic cycle of ≈ 22 years (the Hale cycle), associated with the reversal of the global magnetic-field polarity (Svalgaard et al. 1978; Usoskin 2017). The strength of the polar field around solar minimum and its subsequent reversal are closely related to the amplitude and evolution of the following cycle.

Solar Cycle 25 has been the subject of multiple predictions regarding its amplitude and timing of maximum (Pesnell 2022; Penza et al. 2023). Recent observations point to a more active cycle than initially expected, with SSN, F10.7, and eruptive activity reaching values comparable to or higher than those of Cycle 24. Observationally constraining the interval of maximum and characterizing its properties is essential to assess the performance of cycle-prediction models and to support space-weather risk mitigation.

Furthermore, solar differential rotation plays a central role in the generation and evolution of the magnetic field through the solar dynamo (Charbonneau 2010, 2020). Rotation is faster at the equator than at higher latitudes and depends on depth, as demonstrated by helioseismology (Christensen-Dalsgaard et al. 1996). Characterizing rotation, both sidereal and synodic, and its variation with latitude provides key information on the coupling between the solar interior and atmosphere.

The present article is a concise version of a more extensive analysis presented in the Master's thesis "Dinámica solar del ciclo 25 a partir de parámetros observacionales y estadísticos" (Dávila Gutiérrez and Mosquera Hadatty 2025), available on Zenodo. That thesis details the numerical methodologies, data processing, and the complete set of results.

The aims of this work are twofold: (i) to determine the time interval of the maximum activity of Cycle 25 from the combined analysis of SSN, Wolf number, F10.7 flux, CMEs, flares, and polar field, and (ii) to estimate solar differential rotation during this cycle from a dedicated observing campaign and to compare the results with SDO and SWPC–NOAA data.

This paper provides two specific contributions beyond previous studies of Solar Cycle 25. First, it introduces a percentile-based F10.7 activity scale (EPH) that allows the maximum-activity interval to be identified in a statistically robust and instrument-independent way. Second, it combines this global statistical framework with an independent ground-based rotation campaign around the activity maximum, thus linking large-scale cycle diagnostics with local measurements of photospheric and chromospheric rotation. Together, these aspects offer a compact framework for benchmarking future cycle-prediction schemes for Cycle 26 and beyond.

## 2. Data and Methods

### 2.1. Solar Activity Indices and Polar Field

To characterize the activity of Cycle 25 we use the following indicators:

– Daily and monthly sunspot number (SSN) from SILSO.
– Monthly Wolf number, defined as
$$R = \kappa (10 g + s), \quad (1)$$

where g is the number of sunspot groups, s is the number of individual spots, and $\kappa$ is a correction factor (Usoskin 2017). Equation (1) provides the classical Wolf index used as a long-term activity tracer.
– Solar 10.7 cm radio flux (F10.7) in solar flux units (sfu), obtained from NOAA/SWPC. This flux correlates with UV/EUV radiation and coronal activity (Tapping 2013; Chen 2011).
– CME occurrence rate and speed distribution, from operational catalogues.
– X-ray flares (classes C, M, and X) recorded by GOES.
– Polar magnetic field from WSO, which provides the evolution of the global polarity since 1974.

The polar field is used as a long-term magnetic proxy: the change of sign of the northern and southern poles marks the transition in the Hale cycle and typically occurs in the vicinity of the maximum of the 11-year cycle.

### 2.2. Dedicated Observing Campaign

The observing campaign was carried out in San José, Costa Rica, on 8, 9, and 10 April 2025. Image sequences were obtained in two regimes:

– Photosphere (white light).
– Chromosphere (Hα).

Each day, two observing blocks per spectral band were scheduled, separated by ≈ 20 minutes, in order to capture the displacement of photospheric and chromospheric tracers. The sequences were recorded as videos, from which stacked images with high signal-to-noise ratio were produced. The resulting

FITS images were used as the basis for geometric and photometric measurements. The detailed observing plan (times, filters, configurations) is described in Dávila Gutiérrez and Mosquera Hadatty (2025).

## 2.3. Instrumentation

For the photosphere, we used white-light telescopes equipped with appropriate solar filters and motorized mounts capable of tracking the diurnal motion with sufficient stability to obtain high-quality video sequences.

For the chromosphere, we employed a double-stack Coronado telescope with an Hα filter centered at 656.3 nm, enabling observations of prominences, filaments, plages, and fine chromospheric structure (Athay 1976; Beckers 1964; Bjørgen et al. 2017). Aperture, focal length, mount, and camera characteristics are detailed in the thesis and summarized, where necessary, in the corresponding tables.

## 2.4. External Data Sets

The statistical analysis of solar activity is supported by the following data sets:

– SILSO: international sunspot number (ISN) series with daily and monthly data.
– NOAA/SWPC: F10.7 flux, X-ray flares, CMEs, and geomagnetic indices.
– WSO: polar magnetic field since 1974.
– SDO/SOHO (NASA): images and magnetograms of the photosphere, chromosphere, and corona.
– OMNIWeb: solar-wind parameters (used in a complementary manner in the thesis).

## 2.5. Image Processing

Images were obtained from stacked frames selected from each video, optimizing signal-to-noise ratio and sharpness. The basic workflow was:

1. Selection of frames with the best seeing conditions.
2. Stacking and alignment to produce a high-quality image.
3. Conversion to FITS format for quantitative analysis.
4. Determination of the solar disk centre and radius.
5. Orientation correction to place solar north upward, using the B0, P, and L0 parameters.
6. Transformation from pixel coordinates to heliographic coordinates (longitude, latitude).

The full geometric procedures are described in detail in Dávila Gutiérrez and Mosquera Hadatty (2025).

## 2.6. Construction of the F10.7–EPH Scale

We used the adjusted F10.7 series from Penticton (Canada) spanning 2004–2025. The procedure was:

1. Compute the monthly mean of the adjusted F10.7 flux.
2. Construct the distribution of monthly values over the full interval.
3. Compute the percentiles (5, 25, 60, 90, 100).
4. Define activity levels (EPH) associated with F10.7 ranges, from low to extreme activity.

The resulting scale, based on this percentile statistics, is summarized in Table 1 and provides a quantitative classification of F10.7 activity levels. Because the percentile boundaries are derived from the 2004–2025 distribution, which spans Solar Cycles 23, 24, and the ascending and maximum phases of 25, the EPH levels are anchored in the long-term behaviour of F10.7 and are therefore robust against cycle-to-cycle amplitude changes. This makes the EPH scale directly reusable in space-weather applications that require a compact classification of radio-flux activity levels.

| HD–P level | Adjusted F10.7 range (sfu) | Percentile | Physical interpretation |
|---|---|---|---|
| 0 | < 67.8 | P0–P5 | Deep minimum. This is the absolute background level of the solar cycle. |
| 1 | 67.8–73.1 | P5–P25 | Low activity. Marks the onset of the ascending or declining phase. |
| 2 | 73.1–103.1 | P25–P60 | Moderate activity. Corresponds to the growth or intermediate phase of the cycle. |
| 3 | 103.1–154.5 | P60–P90 | High activity. Represents the transition toward solar maximum. |
| 4 | > 154.5 | P90–P100 | Solar maximum. Emission is intense and disturbances are significant. |

Table 1. EPH scale: Classification of the F10.7 solar radio flux. Source: authors' own work.

In the following sections, the EPH scale is used as a central tool to identify and quantify the maximum-activity interval of Solar Cycle 25.

## 2.7. Estimation of Rotation Velocities

Solar rotation is differential, with an angular velocity that decreases from the equator toward higher latitudes. An empirical expression widely used to describe this latitudinal dependence is (Newton and Nunn 1951; Balthasar et al. 1986; Beck 2000):

$$\omega(\varphi) = A + B \sin^2\varphi + C \sin^4\varphi, \quad (2)$$

where $\varphi$ is the heliographic latitude, A is the equatorial velocity, and B and C describe the decrease toward higher latitudes. Equation (2) summarizes the classical form of the surface differential-rotation law, which we use mainly as a reference to compare the measured values with classical empirical relations.

Rotation was estimated from three data sources:

a) Our own images: tracking of sunspots and active regions between pairs of images separated by ≈ 20 minutes.
b) SDO images: measurement of linear displacements (km) of active regions between successive images and conversion to linear (km h$^{-1}$) and angular velocities, given the latitude.
c) SWPC–NOAA data: use of the reported heliographic longitudes of active regions on consecutive days.

In all cases, the synodic angular velocity was computed as

$$\omega\_syn = \Delta\lambda / \Delta t, \quad (3)$$

where $\Delta\lambda$ is the longitudinal displacement in degrees and $\Delta t$ is the time interval in days. Equation (3) is applied systematically to obtain $\omega\_syn$ from the observed tracer displacement.

The sidereal and synodic periods are related to the angular velocity by (Newton and Nunn 1951; Beck 2000):

$$P\_sid = 360 / (\omega\_syn + 0.9856), \quad (4)$$
$$P\_syn = 360 / (\omega\_sid - 0.9856), \quad (5)$$

where 0.9856° day$^{-1}$ is the mean angular velocity of the Earth. Equations (4) and (5) are used throughout this work to convert between sidereal and synodic periods from the measured angular velocities.

The results of the rotation measurements obtained from our own data and from SWPC–NOAA are presented in Tables 3 and 4 and summarized in Fig. 4, which shows the dependence of the angular velocity on latitude.

## 2.8. Uncertainties

We considered the following sources of error:

– Determination of the solar disk centre.
– Orientation of solar north.
– Spatial resolution of the images.

– Accuracy of the time stamps of the observations.

Error propagation was applied to the angular velocities obtained from Eq. (3) and to the derived periods obtained from Eqs. (4) and (5). For external data sets, the reported uncertainties or those inherent to the sampling were taken into account, as discussed in more detail in Dávila Gutiérrez and Mosquera Hadatty (2025).

## 3. Results

## 3.1. Maximum of Solar Cycle 25

### 3.1.1. SSN and Wolf Number

Figure 1 shows the daily sunspot number between 2018 and 2025. A sustained increase in activity is observed from 2022 onward, with daily peaks exceeding 200 spots in the second half of 2024. The monthly SSN reaches its maximum of 216 in August 2024, the highest value of Cycle 25.

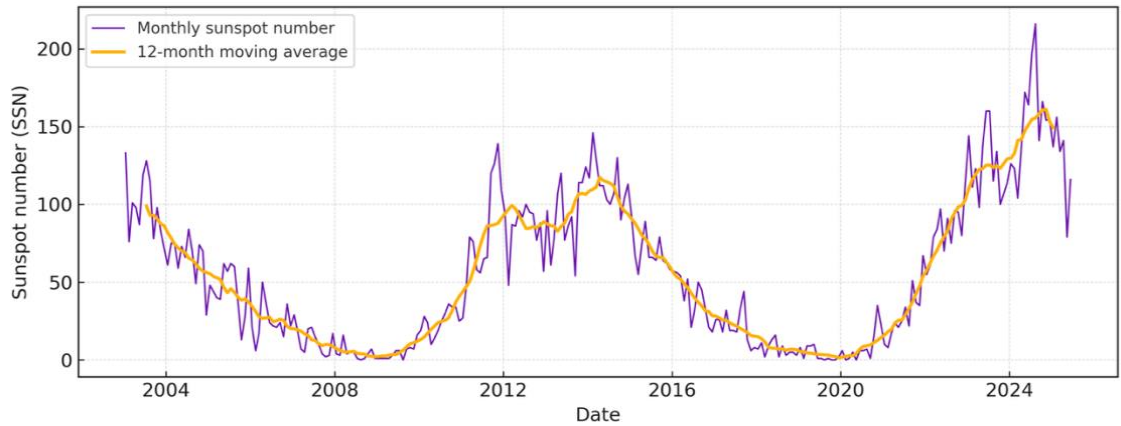

Figure 1. Daily sunspot number between the years 2004 and 2025. Source: author own work based on daily relative sunspot number (SN) data from the SILSO observatory.

The Wolf number, defined by Eq. (1), exhibits a consistent behaviour, with monthly values exceeding 200 during 2024–2025. These indicators identify 2024–2025 as the phase of maximum activity, in agreement with previous prediction studies (Pesnell 2022; Penza et al. 2023).

### 3.1.2. F10.7 Flux and EPH Scale

The adjusted F10.7 series shows a sustained increase from 2022 onwards, with progressively higher daily and monthly values. The EPH scale defined in Sect. 2.6 allows these values to be classified quantitatively into activity levels, from low to extreme.

Figure 2 shows the distribution of F10.7 in terms of percentiles and the boundaries of the EPH scale, indicating that the values corresponding to very high and extreme activity levels (EPH $\geq$ 3) are mainly concentrated between August 2024 and January 2025, with an extension into the first quarter of 2025.

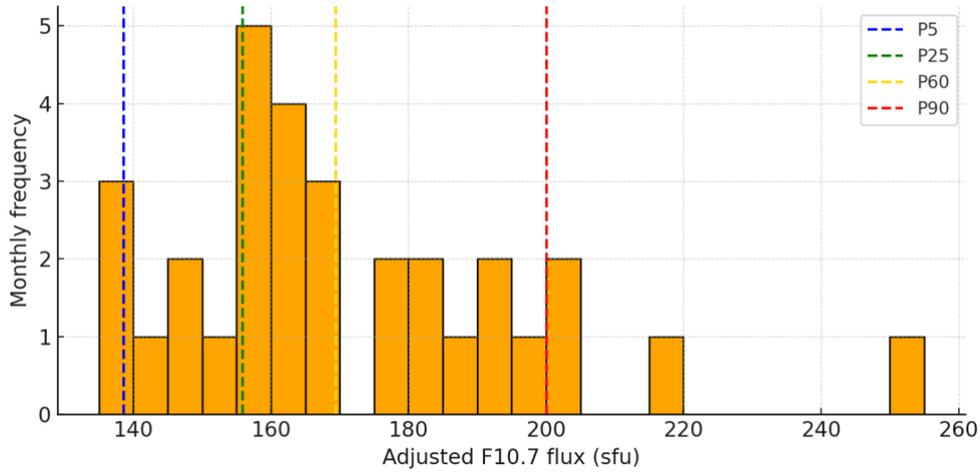

Figure 2. Distribution of the F10.7 solar flux by percentiles. Source: authors' own work based on daily data from Space Weather Canada.

Because the EPH boundaries are defined from the 2004–2025 distribution, which includes the late phase of Cycle 23, Cycle 24, and the rise and maximum of Cycle 25, the scale is anchored to the long-term statistical behaviour of F10.7. In practice, the EPH scheme can be used as a simple "activity flag" for operational purposes: for example, EPH ≥ 3 systematically marks intervals with enhanced CME and flare rates during Cycle 25.

### 3.1.3. CMEs and Flares

The analysis of CME catalogues since 1997 shows a clear increase in the event rate during Cycle 25, with several fast CMEs (> 2000 km s$^{-1}$) in the interval from August 2024 to January 2025.

Regarding flares, more than 190 M-class and 19 X-class events were recorded in Cycle 25, with the X6.3 flare of 10 December 2024 being particularly noteworthy. The temporal clustering of these events is consistent with the maximum-activity interval defined by SSN, Wolf number, and F10.7.

### 3.1.4. Polar Magnetic Field

Figure 3, based on WSO data, shows the evolution of the solar polar magnetic field from 1974 to July 2025. For Cycle 25, the reversal of the northern pole is observed on 1 August 2024 and of the southern pole on 4 January 2025.

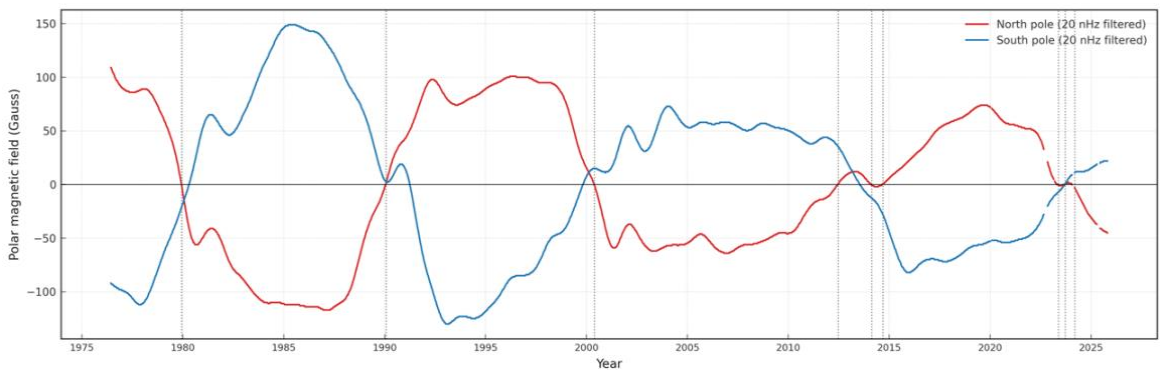

Figure 3. Magnetic polar reversal of the Sun. Evolution of the solar polar magnetic field from 1974 to 31 July 2025. Source: authors' own work based on data from the Wilcox Solar Observatory (WSO), Stanford University.

These dates coincide with the maximum-activity interval defined by the combination of SSN, Wolf number, F10.7, and eruptive activity, reinforcing the identification of the period August 2024–January 2025 as the maximum of Cycle 25.

**3.1.5. Synthesis of Indicators**

Table 2 summarizes the main indicators of the maximum:

– Peak monthly SSN in August 2024 (216).
– Elevated Wolf numbers during 2024–2025.
– F10.7 values in the upper levels of the EPH scale.
– Increased occurrence of CMEs and M- and X-class flares.
– Polar-field reversal between August 2024 and January 2025.
– Chromospheric active-region index maximum during the same interval.

| Indicator | Source | Observational evidence | Interval of maximum activity |
|---|---|---|---|
| Relative sunspot number (SN) | SILSO | Monthly maximum value of 216 in August 2024; daily peaks above 200 around mid-2024. | August 2024 |
| Wolf number | SILSO | Evolution consistent with SN; elevated monthly values in July (196.8) and August (216) 2024. | July–August 2024 |
| F10.7 cm radio flux | Penticton (Canada) | Mean values above 180 sfu between August 2024 and January 2025. | May 2024–February 2025 |
| Polar magnetic-field reversal | Wilcox Solar Observatory | Northern-hemisphere field reversed in August 2024 and southern hemisphere in January 2025. | August 2024–January 2025 |
| M/X-class flares | NOAA–SWPC | Highest number of energetic events between May and December 2024. | May–September 2024 |
| Chromospheric active-region index (Ca II K line) | Own observations | Maximum chromospheric coverage recorded between August 2024 and February 2025. | August 2024–February 2025 |

Table 2. Observational indicators confirming the solar maximum of Cycle 25. Source: authors' own work.

Taken together, these data show that Cycle 25 has been more active than initially expected and that its maximum was concentrated in the interval August 2024–January 2025.

**3.2. Solar Rotation**

The rotation analysis presented here is intentionally restricted to a short interval around the activity maximum and to a limited set of low-latitude active regions. It is therefore designed as a case study that illustrates how ground-based tracking and space-based data can be combined to obtain consistent rotation measurements, rather than as a full-cycle rotation survey.

The rotation results are summarized in Tables 3 and 4 and in Fig. 4.

From the images obtained in our dedicated observing campaign we derived mean synodic angular velocities of 12.8 ± 2.8° day$^{-1}$ for the northern hemisphere and 13.38 ± 0.54° day$^{-1}$ for the southern hemisphere. Using Eqs. (4) and (5), these values yield sidereal periods of 22.78 days (North) and 21.77 days (South), consistent with classical values for low latitudes (Newton and Nunn 1951; Balthasar et al. 1986; Beck 2000). Table 3 lists the velocities and periods derived from the observing campaign.

| Identification (NOAA) | Rotation rate Ω (deg day⁻¹) | Synodic period (days) | Hemisphere |
|---|---|---|---|
| NOAA14055_1 | 12.7 ± 2.9 | 28.5 ± 3.1 | North |
| NOAA14054_1 | 13.38 ± 0.54 | 26.9 ± 1.1 | South |
| NOAA14056 | 13.36 ± 0.28 | 26.95 ± 0.56 | South |

Table 3. Mean angular rotation rate of the identified regions from the observing campaign. Source: authors' own work based on data from the observing campaign.

The linear velocities derived from SDO images ranged from ≈ 872 km h⁻¹ to more than 7400 km h⁻¹, depending on latitude and date, once the measured linear displacements were converted to angular velocities with Eq. (3) and then to linear speeds using the local solar radius at the corresponding latitude.

SWPC–NOAA data provided synodic angular velocities of ≈ 13–16° day⁻¹ and sidereal periods of 21.19–25.74 days during 8–10 April 2025, again using Eqs. (3)–(5). Table 4 summarizes the angular velocities and the synodic and sidereal periods obtained for the active regions analysed.

| ID | Hemisphere | ω_sin_8–9 (deg day⁻¹) | ω_sin_9–10 (deg day⁻¹) | ω_sid_8–9 (deg day⁻¹) | ω_sid_9–10 (deg day⁻¹) | P_sin_8–9 (d) | P_sin_9–10 (d) | P_sid_8–9 (d) | P_sid_9–10 (d) |
|---|---|---|---|---|---|---|---|---|---|
| NOAA 4046 | North | 13.0 | 16.0 | 13.99 | 16.99 | 27.69 | 25.74 | 22.5 | 21.19 |
| NOAA 4055 | North | 14.0 | 15.0 | 14.99 | 15.99 | 25.71 | 24.02 | 24.0 | 22.52 |
| NOAA 4056 | North | 13.0 | 14.0 | 13.99 | 14.99 | 27.69 | 25.74 | 25.71 | 24.02 |
| NOAA 4057 | North | 13.0 | 14.0 | 13.99 | 14.99 | 27.69 | 25.74 | 25.71 | 24.02 |
| NOAA 4048 | South | 14.0 | 13.0 | 14.99 | 13.99 | 25.71 | 27.69 | 24.02 | 25.74 |
| NOAA 4049 | South | 13.0 | Not visible | 13.99 | Not visible | 27.69 | 25.74 | Not visible | Not visible |
| NOAA 4054 | South | 14.0 | 14.0 | 14.99 | 14.99 | 25.71 | 24.02 | 25.71 | 24.02 |

Table 4. Synodic and sidereal angular rotation rates and periods of active regions from SWPC–NOAA for 8–9 April 2025. Authors' own work based on data from the NOAA Space Weather Prediction Center.

Figure 4 shows the mean angular velocity as a function of latitude for the dates analysed, condensing the information from the different data sources. A decreasing trend in velocity from the equator toward higher latitudes is evident, characteristic of solar differential rotation, together with a slightly higher velocity in the southern hemisphere during the interval studied.

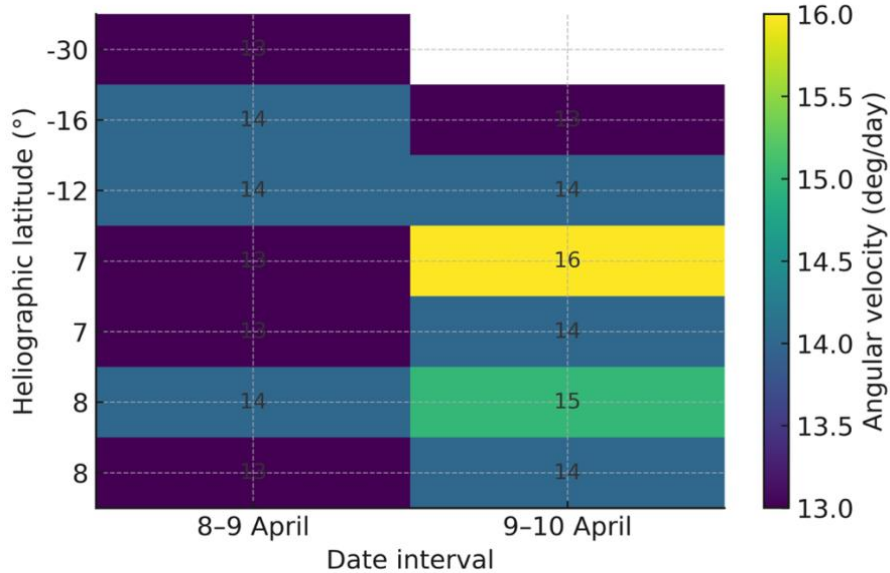

Figure 4. Mean angular velocity as a function of latitude and observation date. Source: authors' own work based on NOAA data.

## 4. Discussion

### 4.1. Differential Rotation and Comparison with the Literature

The sidereal periods obtained (≈ 22–23 days at low latitudes), derived using Eqs. (3)–(5), and the observed latitudinal gradient are consistent with empirical models of differential rotation based on sunspot groups and helioseismic techniques (Newton and Nunn 1951; Balthasar et al. 1986; Beck 2000; Pulkkinen and Tuominen 1998). Although the temporal baseline and sample size are limited, the agreement with empirical rotation laws and with SWPC–NOAA/SDO measurements indicates that the methodology is sound and can be extended to longer time spans in future work.

The agreement between the velocities derived from our own data, SDO, and SWPC–NOAA suggests that tracer tracking on calibrated images is an adequate methodology to measure solar rotation, even with privately owned instrumentation, provided that the geometric processing is carried out carefully.

### 4.2. Cycle 25 Maximum and Predictions

The August 2024–January 2025 interval defined here is consistent with recent predictions for the maximum of Solar Cycle 25 (Pesnell 2022; Penza et al. 2023) and with the evolution of the polar field (Svalgaard et al. 1978) and F10.7 (Tapping 2013). The more active behaviour than initially expected reinforces the need to update cycle models regularly, incorporating high-quality data and statistical approaches such as the proposed EPH scale.

The combination of indicators (SSN, Wolf number, F10.7, CMEs, flares, polar field) provides a robust framework for determining the maximum, both from the perspective of solar physics and of space weather. In this sense, the EPH scale and the combined use of global indices and targeted rotation measurements offer a compact framework to benchmark future cycle–prediction schemes for Cycle 26 and beyond.

### 4.3. Methodological and Observational Contribution

This study shows that an observing campaign conducted from non-institutional facilities can provide useful rotation measurements and support the identification of the cycle maximum, as long as it is integrated with international databases.

The percentile-based quantitative scale for F10.7 complements traditional indicators and facilitates the classification of activity levels in studies of solar cycles and space weather. Because it is anchored to the long-term distribution of F10.7, the EPH scheme can be used as a simple activity flag in operational contexts. For example, EPH ≥ 3 systematically identifies intervals with enhanced CME and flare rates during Cycle 25 and can be used in future work to select epochs of interest for detailed space-weather studies. A more detailed description of this methodology and its applications can be found in Dávila Gutiérrez and Mosquera Hadatty (2025).

### 4.4. Limitations

The main limitations arise from the temporal and spatial resolution of our own observations, the selection of tracers, and the dependence on external catalogues with their own processing criteria. Nevertheless, the consistency among the different data sets supports the robustness of the overall results. Future extensions of this work could include longer time spans for rotation measurements and the application of the EPH scale to interplanetary and geomagnetic indices.

## 5. Summary and Conclusions

In this article we analysed Solar Cycle 25 combining global activity indices, eruptive phenomena, and a dedicated observing campaign. Our main conclusions are:

1. The combined analysis of SSN, Wolf number, F10.7, CMEs, flares, and polar field indicates that Cycle 25 reached its maximum activity between August 2024 and January 2025.
2. The reversal of the polar magnetic field, with the northern pole changing sign on 1 August 2024 and the southern pole on 4 January 2025, coincides with this interval, providing additional confirmation of the maximum.
3. Our observations from April 2025 yield mean synodic angular velocities of $12.8 \pm 2.8°$ day$^{-1}$ (northern hemisphere) and $13.38 \pm 0.54°$ day$^{-1}$ (southern hemisphere), corresponding to sidereal periods of 22.78 and 21.77 days, respectively, in agreement with the literature.
4. The linear velocities derived from SDO (≈ 872–7400 km h$^{-1}$) and the angular velocities obtained from SWPC–NOAA (13–16° day$^{-1}$; periods of 21.19–25.74 days) confirm differential rotation and show a slightly higher velocity in the southern hemisphere during the analysed interval.
5. The percentile-based methodology applied to the F10.7 flux (EPH scale) is useful for identifying intervals of maximum activity and can be employed in future studies of solar cycles and space weather. Because it is derived from a multi-cycle distribution, it is robust against amplitude changes and instrumental differences.
6. The combination of ground-based observations with space-borne data sets demonstrates the potential of collaborations between non-institutional observers and the professional heliophysics community, and provides a framework to benchmark future predictions for Cycle 26.

**Acknowledgements**

I acknowledge the academic guidance received during the development of the Master's thesis, as well as access to data from NASA, SILSO, NOAA/SWPC, WSO, and the SDO and SOHO missions. We also thank those who contributed to the observing campaign and the support of family and friends.

**Funding Information**

No specific funding was received for this study.

**Data Availability**

The data sets analysed in this study are publicly available from the corresponding repositories: SILSO sunspot number (Royal Observatory of Belgium), NOAA/SWPC (F10.7, flares, CMEs), WSO polar fields (Stanford University), and SDO/SOHO imagery and magnetograms (NASA). Details of their use are provided in the text and in Dávila Gutiérrez and Mosquera Hadatty (2025).

**Competing Interests**

The author declare that they have no competing interests.